\documentstyle[osa,manuscript,graphicx]{revtex}

\begin{document}

\begin{center}{\Large \bf Probing the Nature of Compactification
with Kaluza-Klein Excitations at the Large Hadron Collider} \\
\vskip.25in
{Pran Nath$^a$, Youichi Yamada$^b$, and Masahiro Yamaguchi$^b$ }

{\it
a Department of Physics, Northeastern University, Boston, 
MA 02115-5005, USA\\
b Department of Physics, Tohoku University, Sendai 980-8578, Japan
} \\%

\end{center}

\begin{abstract}                
It is shown that the nature of compactification of extra dimensions
in theories of large radius compactification can be explored 
in several processes at the Large Hadron Collider (LHC). Specifically it 
is shown
that the characteristics of the Kaluza-Klein (KK) excitations encode 
information on the nature of compactification,
i.e., on the number of compactified dimensions as well as on the
type of compactification, e.g., of the specific orbifold compactification. 
The most dramatic signals arise from the interference pattern involving 
the exchange of the Standard Model
spin 1 bosons  ($\gamma$ and Z) and their Kaluza-Klein modes
in the dilepton final state $pp\rightarrow l^+l^-X$.
It is shown that LHC with 100$fb^{-1}$ of luminosity can discover
Kaluza-Klein modes up to compactification scales of $\approx 6$ TeV 
 as well as identify 
the nature of compactification.
Effects of the Kaluza-Klein excitations of the W boson and of the
gluon are also studied.  
 Exhibition of these phenomena is given 
for the case of one extra dimension and for the case of 
two extra dimensions with $Z_2\times Z_2$,
$Z_3$, and $Z_6$  orbifold compactifications.
\end{abstract}

\newpage


Recent analyses of the effects of extra space time dimensions on the 
Fermi constant and on the precision electro-weak 
data\cite{gf,mp,gmu,marciano} have led to 
a lower bound on the scale of the extra dimensions that 
well exceeds 1 TeV and could be as high as 5 TeV for the case of one
extra dimension ($d=1$),  and the lower bound  is even higher for 
the case $d>1$\cite{gf}. Given these bounds, it appears very unlikely that 
the Kaluza-Klein (KK) excitations associated with the Standard Model
gauge bosons ($\gamma$, $W$, $Z$ and gluon)  can be observed at the upgraded 
Tevatron. However, the possibility exists that these modes could be 
directly produced at the Large Hadron Collider (LHC). In this paper
we explore this possibility.
 
 There is a long and rich history associated with Kaluza-Klein 
 models\cite{kaluza}. More recently the interest in these models 
 has been revived\cite{a,l,a1,d,s,gf,mp,gmu,marciano} because there exists the 
 possibility that models with low scale compactifications might
 arise in string theory, specifically in the context 
 of Type I string and low energy consequences of
 such scenarios have been investigated.
 Our analysis is in the spirit of Ref. 1 where we work within the context
 of an effective field theory of a $p$-brane of a Type I string where
the matter and the gauge fields of the Standard Model reside  while gravity
 propagates in all 10 dimensions. Our analysis corresponds to 
 compactifications internal to the $p$-brane ($p=d+3$ where $d$ is the number
 of extra space time dimensions), and there is no influence
 on the analysis of compactifications which are transverse to the 
 brane where only gravity propagates.  The purpose of this 
 analysis is to investigate the possibility of observing the 
 Kaluza-Klein excitations of the Standard Model gauge bosons 
arising from the compactifications of the extra 
 dimensions. We shall show that several channels provide clear 
 signals for the possible observation of such states. 

 In the analysis below we first consider the Drell-Yan process
  $pp\rightarrow l^+l^-+X$ where the Kaluza-Klein modes of the $Z$-boson 
  as well as of the photon contribute. 
Here we find that the production cross section of a charged lepton pair in the
$d=1$ case is about one order of magnitude larger than that for the
sequential Standard Model $Z'$ boson with the same mass, 
and the enhanced cross section allows one to extend 
the search for the Kaluza-Klein mode significantly. Thus LHC with
100$fb^{-1}$ luminosity will be able to discover these 
Kaluza-Klein modes up to $M_R 
\approx$ 6 TeV.  In the analysis we find a phenomenon specific 
to the Kaluza-Klein models in that the dilepton production cross section
for the process $pp\rightarrow l^+l^- +X$ as a function of the  dilepton
 invariant mass has a dip just
below the resonance peak of the Kaluza-Klein state. The dip arises
from a negative
interference between the exchange of the photon and for the Z boson
and their Kaluza-Klein recurrences.  
 We also study the processes $pp\rightarrow
  l^{\pm} \nu_l+X$ and 
 $pp \rightarrow jj+X$ where the Kaluza-Klein recurrences of the W boson
and those of the gluon  respectively are involved,  
and show that they also provide
interesting signals for the discovery of Kaluza-Klein excitations.   

   Another remarkable result seen is that the signatures at the 
    hadron collider can be used to distinguish among different models
    of compactification. Specifically it is shown that for the case
    $d=2$ one can distinguish via the dileptonic signal not only models
    with different number of compactified dimensions but
    also different types of compactifications, such as the
    $Z_2\times Z_2$ orbifold model from the $Z_3$ and $Z_6$
     orbifold models. 

  The details of the models we consider can be found in Ref. 1 and we discuss 
 here some of its salient features. For the case
 D=5 ($d=1$), the model is described by a Lagrangian whose first few terms
 are exhibited below
  
 \begin{equation}
L_5=-\frac{1}{4} F_{MN}F^{MN}
-(D_MH)^{\dagger}(D^MH)
-\bar\psi\frac{1}{i}\Gamma^{\mu} D_{\mu}\psi-V(H)+..
\end{equation}
where $A_M$ (M=0,1,2,3,4) is the vector potential in 5 dimensions, 
$D_M$=$\partial_M-ig^{(5)}A_M$ is the gauge covariant derivative, 
and $H$ is the Higgs doublet. The potential $V(H)$ is arranged to
allow spontaneous breaking of the electro-weak symmetry
which allows the $W$ and $Z$ bosons in 5D to gain electro-weak masses.
We make the assumption that the gauge and the Higgs fields lie
in the bulk while the quark-lepton generations lie on the 4D wall. 
We compactify the model on $S^1/Z_2$ with a radius of compactification
$R$ (and compactification mass scale $M_R=\frac{1}{R}$) and with the 
quarks and leptons residing on one of the orbifold 
points. After compactification the zero modes of the model 
correspond exactly
to the spectrum of the Standard Model (SM). The relative strength 
of the gauge couplings of the 4D gauge fields and their Kaluza-Klein  modes is
given by\cite{gf}  

\begin{equation}
{\it L_{int}}=g_ij_i^{\mu}(A_{\mu i}+\sqrt 2\sum_{n=1}^{\infty} A_{\mu i}^n)
\end{equation}
where $A_{\mu i}$ are the zero modes and  $A_{\mu i}^n$ are the 
Kaluza-Klein modes. We note that the Kaluza-Klein  modes couple more
strongly by a factor of $\sqrt 2$ than their zero mode counterparts.

Throughout this paper, we assume that the Kaluza-Klein modes decay only into 
 fermion pairs. In supersymmetric theories, they may also decay into
sfermions, which will make the decay width 3/2 times larger. 
The main results of our analysis, however, will not be significantly
affected by the inclusion of these effects.

With these preliminaries we discuss now the main results of the 
analysis. We consider first the 
Drell-Yan processes $pp \rightarrow l^+ l^- +X$. (The  relevant 
formulae can be found in Ref.\cite{aguila}).
   In our computations, we used CTEQ5L parton distribution 
   functions\cite{CTEQ}.
In Fig. 1, a plot of the production cross section of a charged lepton pair
from Kaluza-Klein excitations of $Z$ and $\gamma$ as a function of the 
compactification scale $M_R$ is given. The plot is for one generation of 
leptons, and no summation over generations is taken.  
For comparison, we also
plot the same production cross section for the sequential Standard
Model (SSM) $Z'$ boson which has exactly the same couplings with the quarks
and leptons as the $Z$ boson does.  We find that the cross section in the
Kaluza-Klein  case is about one order of magnitude larger than that 
for the case of the SSM $Z'$ boson. This result arises in part because as noted earlier 
 the coupling of the Kaluza-Klein gauge boson is enhanced by $\sqrt{2}$ 
 and in part  
 because there is a constructive interference between the photonic 
 Kaluza-Klein mode and the Z  boson Kaluza-Klein mode which lie close to
 each other and essentially overlap.  In Ref.\cite{rizzo},
 the discovery reach for the sequential Standard Model $Z'$ boson at the LHC 
with  a luminosity of 100 
$fb^{-1}$ (5 events for one-generation lepton pair) was found to be 5 TeV.  
We find that the corresponding reach for the 
Kaluza-Klein  mode will give  
$M_R$ around 6 TeV. We note that this 
mass region is not yet excluded by the  
constraint from the electroweak precision data\cite{gf,mp,gmu,marciano}.

In Fig. 2 the cross section $d\sigma/dm_{ll}$ is given as a function of
the dilepton invariant mass $m_{ll}$ for the cases where 
$M_R=2, 5, 8$ TeV.  For comparison a plot of  the
contribution from only the Standard Model $Z$ boson and the photon
exchange is also given. The plot exhibits  clear resonance 
peaks corresponding to the masses of the Kaluza-Klein  states. 
We note that each peak
is a superposition of the contribution from the exchange of a Kaluza-Klein
excitation  of the photon and from the exchange of a Kaluza-Klein 
excitation of the $Z$ boson. A close scrutiny shows that the resonances
are not of the typical Breit-Wigner form due to the fact as mentioned 
above  that each 
resonance is a superposition of  two resonances one from the
Kaluza-Klein excitation of the photon and the other from the Kaluza-Klein
excitation of the Z boson. The superposition of the two resonances which
have significantly different widths gives a  composite resonance 
 shape which is a distorted Breit-Wigner form. However, it remains to be 
seen if the LHC detectors will be sensitive enough to observe such 
distortions. 
 A more interesting phenomenon arises  from the 
 negative interference between the exchange contributions of the 
 gauge bosons and of their Kaluza-Klein excited states below the 
 peak of the Kaluza-Klein excitation. The negative
 interference leads to a sharp dip below the peak. This signal
  is specific to  the Kaluza-Klein excitations
 and would not arise for any other type of excitation which are
 not recurrences of the Standard Model gauge
 bosons. One may worry if the loop corrections to the gauge coupling 
 constants of the gauge boson and their Kaluza-Klein modes would 
 diminish or even eliminate the negative interference effect. 
 However, the loop corrections are expected to be small and  we have 
 checked that the appearance of the dip is stable against  small
  perturbations  of the gauge couplings. 

We next consider $pp \rightarrow l^{\pm} \nu_l +X$ through the exchange of 
the Kaluza-Klein excitations of the $W$ boson. Plot of the production 
cross section is given in Fig. 3
as a function of $M_R$ for the case $d=1$. 
Compared to the case of the sequential Standard Model $W'$ boson exchange
which is also shown in Fig. 3 one
finds an enhancement of the cross section for the 
Kaluza-Klein case. 
Another interesting process concerns the dijet production 
$pp \rightarrow j j +X$ ($j\neq t$)
via the exchange of the gluon and its Kaluza-Klein excitation.\footnote{
It is interesting to mention that the flavor universal coloron model shares 
similar properties with the KK modes of the gluon\cite{coloron}.}  
In Fig. 4 a plot of $d\sigma/dm_{jj}$ is given as a function of the 
dijet invariant mass $m_{jj}$. In this process t-channel as
well as s-channel exchanges contribute.  Further, the resonance widths
 are much broader than for the dilepton case of Fig. 2.  
 Because of the combination of these two effects the resonance peaks 
 are not so manifest for this process as is evident from Fig. 4.
  However, a signal for the Kaluza-Klein
 excitations still exists since  one finds an excess of
the dijet events with large invariant mass when compared to the Standard Model
case.  This is evident in Fig. 5 where  the production cross section 
of the two jets with invariant mass larger than 1, 3, and 5 TeV is 
shown and the cross section for the last two cases  exceed
that of the SM case (corresponding to $M_{R} \rightarrow \infty$). Thus
we conclude that dijet production is also a promising channel to probe 
the extra dimensions. We note that all Standard Model gauge bosons have
common Kaluza-Klein mass spectra.

Finally we consider the case of more than one extra dimension ($d>1$).
Here the information regarding the number of extra dimensions may
appear directly in the decay pattern of the Kaluza-Klein excitations.
For illustration we consider the case of d extra dimensions with the
compactification $S^1/Z_2\times S^1/Z_2\times ..\times S^1/Z_2$, i.e., each
extra dimension is compactified on a $S^1/Z_2$ and we assume that each
circle has a common radius $R$. In this case there are $d$ degenerate
states at the first Kaluza-Klein excitations, each of which has a mass
$M_R$ and a coupling enhancement factor of $\sqrt{2}$ just as in the $d=1$
case. Although the decay width of each Kaluza-Klein mode at the
lowest level is the same as the one in $d=1$, the multiplicity
enhances the cross section around the resonance peak by $d^2$ compared
to the $d=1$ case. Thus the height of the resonance peak at the first
Kaluza-Klein excitation, for example, provides a direct count of the number 
of the extra dimensions.   For the
second Kaluza-Klein excitation, there are $d(d-1)/2$ states with
mass $\sqrt{2}M_R$ and a coupling enhancement factor of $2$.  
Similar analyses can
 be done for the higher Kaluza-Klein states and the dependence on the
 number of extra dimensions identified.
  
 Unlike the $d=1$ case the compactifications for $d>1$ 
 are much more model dependent. 
 To illustrate this point we consider the $d=2$ case in detail 
and compare the following models of compactification: (1) a $Z_2 \times Z_2$
orbifold model as above where the compactified space is 
$S^1/Z_2 \times S^1/Z_2$ 
and the two $S^1$ are assumed to have the common radius $R$, and (2)
 $Z_3$ and $Z_6$ compactifications with a two-dimensional torus of periodicity
$2\pi R$.  The analysis of case (1) is given in Table1 where 
we list the masses, the multiplicities, and the 
coupling enhancement of the Kaluza-Klein vector bosons to the 
boundary fermions relative to their zero modes. 
The $Z_3$ and $Z_6$ orbifold compactifications have an interesting
relationship. For the $Z_3$ orbifold compactification of 2d the 
mass formula for Kaluza-Klein states is 

\begin{equation}
M^2_{Z_3}=\frac{4}{3R^2}(m_1^2+m_1m_2+m_2^2)
\end{equation}
where $m_1,m_2$ take on positive and negative integer values, while
for $Z_6$ orbifold compactification of 2d the analogous mass formula is

\begin{equation}
M^2_{Z_6}=\frac{4}{3R^2}(m_1^2-m_1m_2+m_2^2)
\end{equation}
Thus one finds that the transposition $(m_1,m_2)\rightarrow (m_1,-m_2)$
takes one from the $Z_3$  mass relation to the $Z_6$ mass relation.
In Table 2 we list the masses, the multiplicities, and the
coupling enhancement of the first few Kaluza-Klein vector bosons to the 
boundary fermions relative to the zero modes for the $Z_3$ and $Z_6$ case.
The analysis of Table 2 shows that for the case of couplings to the 
boundary  fermions the cross sections for the production of Kaluza-Klein
states for $Z_3$ and $Z_6$ are the same. Thus we shall discuss only 
$Z_3$ in detail and similar results hold for the $Z_6$ compacfication.
 A comparison of Table 1 and Table 2
shows that the  masses of the  Kaluza-Klein excitations, their 
multiplicities and the strength of their couplings to the
boundary fermions depend on the nature of compactification. 
These attributes will manifest in the production cross-section and
in the resonance structure of these states at the LHC. An exhibition
of this phenomena is given in Fig. 6 where the cross section for the
process $pp\rightarrow e^+e^- +X$ is given  for the case d=1 and for
the case d=2 for the two orbifold compactifications, $Z_2\times Z_2$ 
and $Z_3$ where the mass of the first Kaluza-Klein excitation is taken to
be 3 TeV for each case. The analysis of Fig. 6 shows that the
three cases can be distinguished by a detailed study of the dileptonic
cross section as a function of the dilepton invariant mass.
We note that a study of this channel can  allow one to
differentiate the case when the radii of compactification are 
unequal. In this case the degeneracy of the Kaluza-Klein states (as in 
case (2) for d=2) will be lifted. However,  
the pattern of the resonance peaks will be more complex resulting
in a richer structure of the dileptonic cross section as a 
function of the dilepton invariant mass. 
We thus conclude that a close study of the resonance structure of the
Kaluza-Klein states will allow one to determine the dimensionality of 
the compactified space as well as the detailed nature of 
compactification.

\begin{center}{\bf Table~1:~ $Z_2 \times Z_2$ orbifold model 
for d=2 with common radius $R$  } \\ 
\vspace{0.3cm}
 \begin{tabular}{|c|c|c|c|c|c|}
 \hline
& 1st KK & 2nd KK & 3rd KK & 4th KK & 5th KK \\ \hline   
mass ($\frac{1}{R}$ unit)   & 1 &  $\sqrt{2}$ & 2 & $\sqrt{5}$ & 2$\sqrt{2}$ 
\\ \hline 
multiplicity     &   2 & 1 & 2 & 2 & 1     \\  \hline  
$\frac{g_{\rm Kaluza-Klein}}{g_{\rm zero~ mode}}$ &$\sqrt{2}$ & 2 & $\sqrt{2}$ & 2 & 2 \\ \hline
\end{tabular} 
\end{center}

\begin{center} 
{\bf Table~2:~ $Z_3$ $\&$ $Z_6$ orbifold models for d=2 } \\
\vspace{0.3cm}
\begin{tabular}{|c|c|c|c|c|c|}
 \hline
                 & 1st KK & 2nd KK & 3rd KK  & 4th KK & 5th KK\\ \hline
mass ($\frac{2}{\sqrt{3}R}$ unit)  &  1 & $\sqrt{3}$  & 2& $\sqrt{7}$ & 3  \\  \hline
multiplicity    &   1            & 1          & 1  & 2 & 1    \\  \hline
$\frac{g_{\rm Kaluza-Klein}}{g_{\rm zero~ mode}}$& $\sqrt{6}$ & $\sqrt{6}$& $\sqrt{6}$ & $\sqrt{6}$ & 
$\sqrt{6}$   \\ \hline
\end{tabular} 
\end{center}

In conclusion 
we have discussed the signatures of the Kaluza-Klein
excitations at the LHC associated with extra dimensions with large radius 
compactifications consistent with the precision electro-weak data.
It is shown that the dileptons from the Drell-Yan process
$pp\rightarrow l^+l^-+X$ exhibit a remarkable dip from an interference
and provide an important signal for the discovery of such states
up to $M_R\approx 6$ TeV. It is found that the processes 
 $pp\rightarrow l^{\pm}\nu_l+X$ and $pp\rightarrow jj+X$ provide further 
 signals for the observation of Kaluza-Klein modes.
 We also discussed the d=2 case and showed
that the dilepton signal can distinguish between the d=1 and the d=2 
cases. Further, for the d=2 case we considered the
orbifold compactifications, $Z_2\times Z_2$, $Z_3$, and $Z_6$ and found 
that the $Z_2\times Z_2$ case can be distinguished from the $Z_3$
and $Z_6$ cases.
Thus if the low scale Kaluza-Klein dimensions exist at the scale
accessible to LHC one can not only determine the  
compactificaion scale $M_R$  but also the number of compactified dimensions
and the nature of compactification itself from a detailed study of the
dileptonic  and other signatures in pp collisions.\\

\noindent
Note added: While this paper was in preparation there appeared a 
paper\cite{adq}
by I. Antoniadis, K. Benakli and M. Quir\'{o}s, which has
 an overlap with some of the topics discussed here. \\

The research of PN was supported in part by the National Science Foundation
grant no. PHY-9602074.
The work of YY  was supported in part by the Grant-in-Aid for Scientific 
Research from the Ministry of Education, Science, Sports, 
and Culture of Japan, No.10740106, and the work of MY by the Grant-in-Aid 
on Priority Area 707 "Supersymmetry and Unified Theory of Elementary
Particles", and by the Grant-in-Aid No.11640246 and No.98270.

\begin{figure}
\begin{center}
\includegraphics[angle=270,width=6.0in]{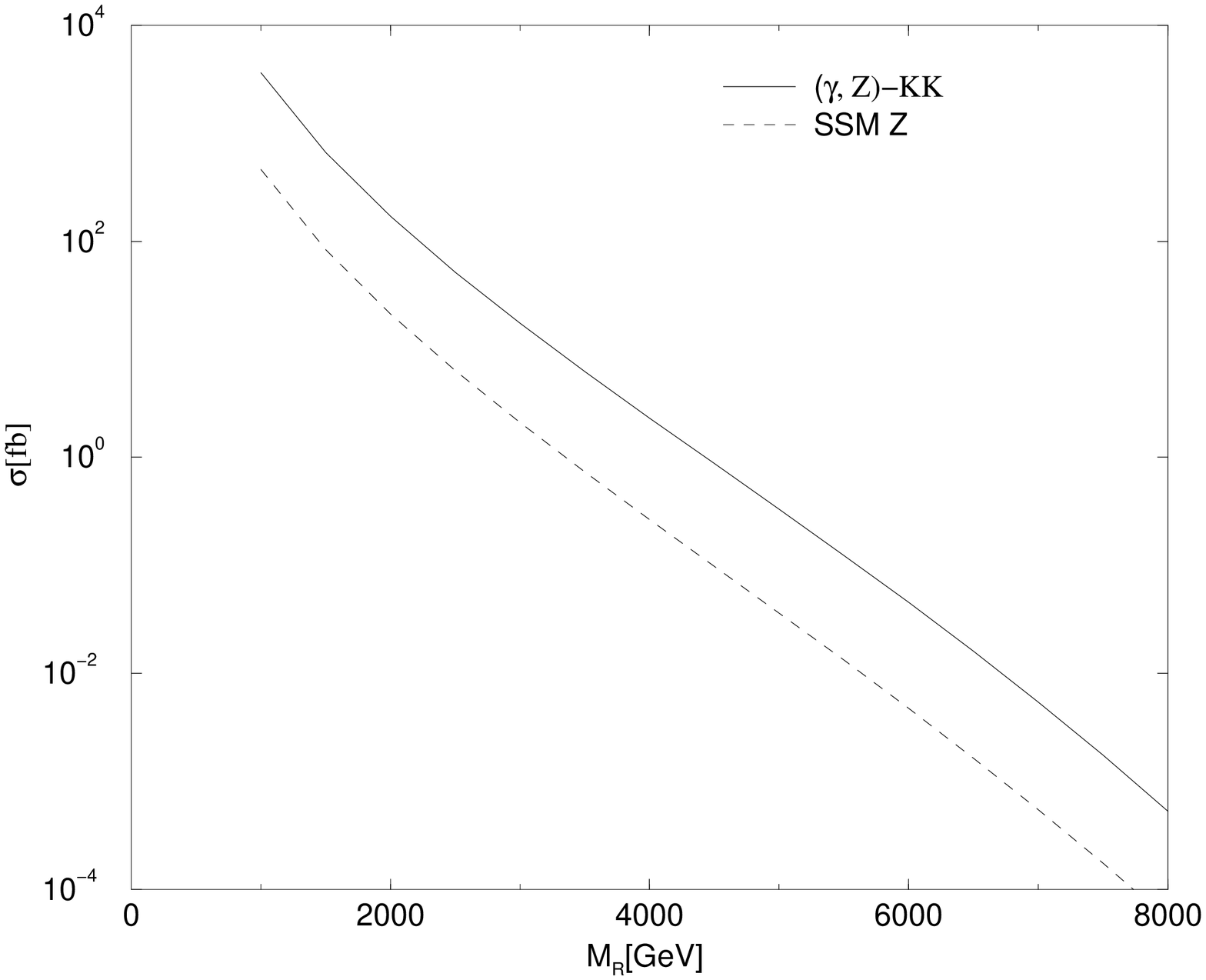}
\caption{Plot of the production cross section of a charged lepton pair
in the process $pp\rightarrow l^+l^-+X$ (solid line) as a function of the 
mass scale $M_R$ via exchange of photonic and Z Kaluza-Klein  modes of the compactified 
dimension. The same analysis for the sequential Standard Model Z' boson
 is shown by the dashed line. } 
\label{fig1}
\end{center}
\end{figure}

\begin{figure}
\begin{center}
\includegraphics[angle=270,width=6.0in]{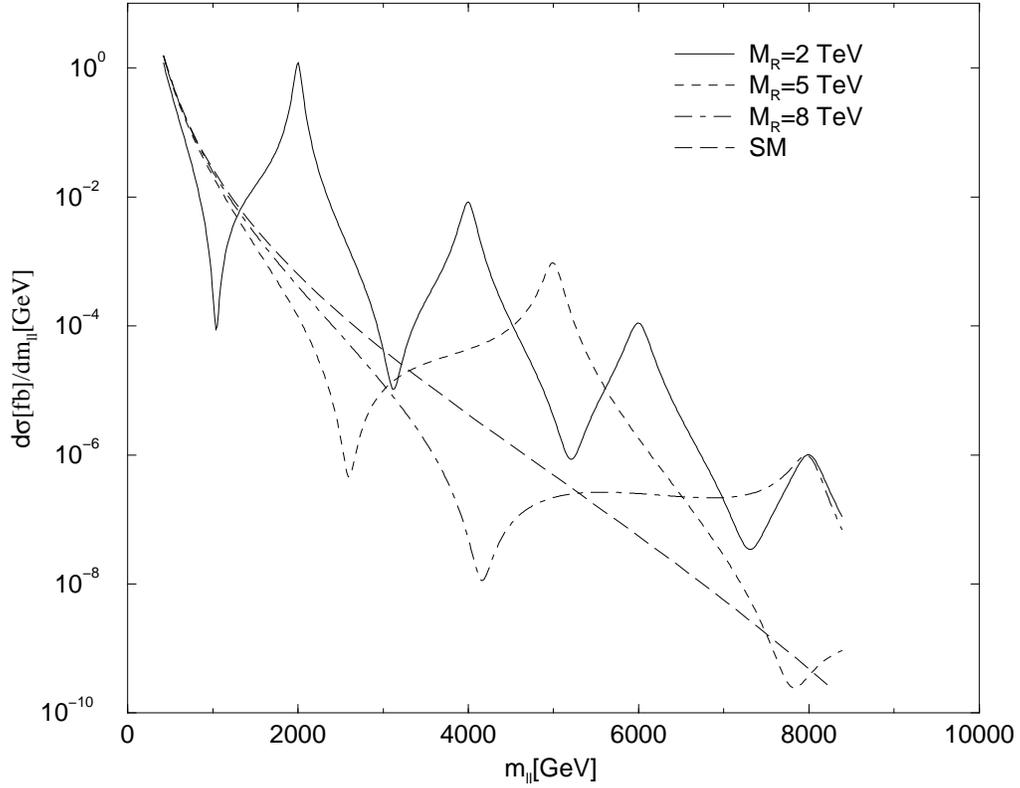}
\caption{Differential cross section $d\sigma/dm_{ll}$ as a function of the
invariant mass $m_{ll}$ of the charged lepton pair for three different
values of the compactified dimension $M_R$. For comparison the analysis
for the SM case is also shown.}
\label{fig2}
\end{center}
\end{figure}

\begin{figure}
\begin{center}
\includegraphics[angle=270,width=6.0in]{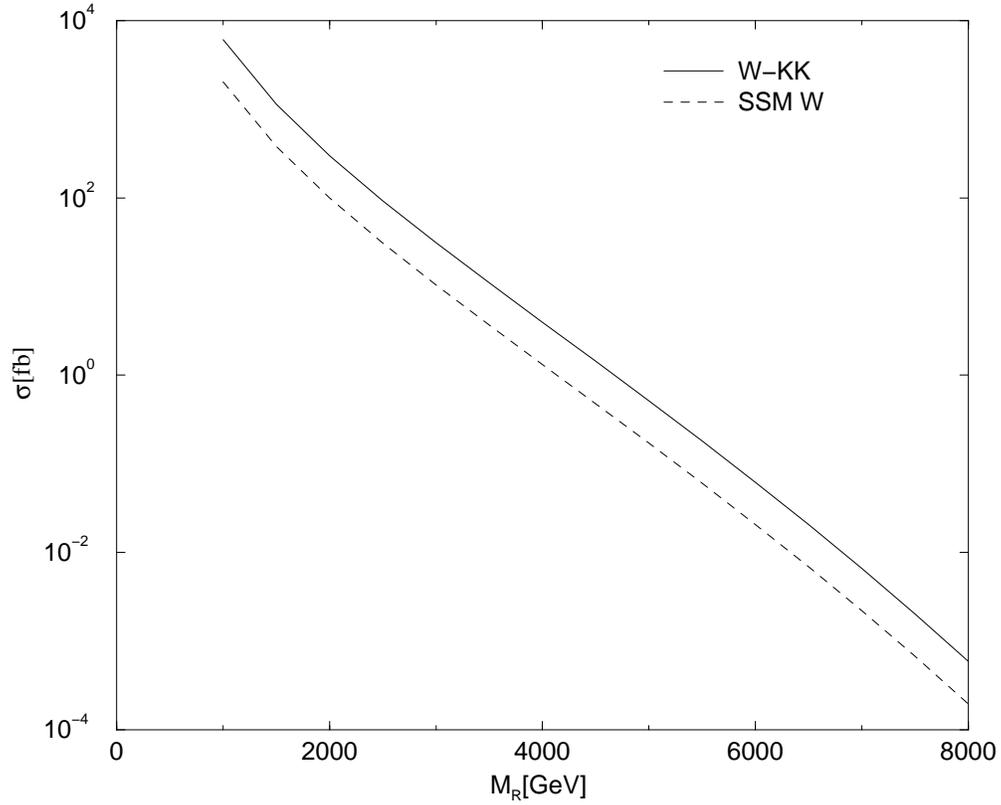}
\caption{Plot of the production cross section $pp\rightarrow l^{\pm}\nu_l +X$
  via exchange of the Kaluza-Klein excitations of $W$ as a function of
  $M_R$ (solid). For comparison the cross section for the sequential
  Standard Model $W'$ boson is also given (dashed).}
\label{fig3}
\end{center}
\end{figure}

\begin{figure}
\begin{center}
\includegraphics[angle=270,width=6.0in]{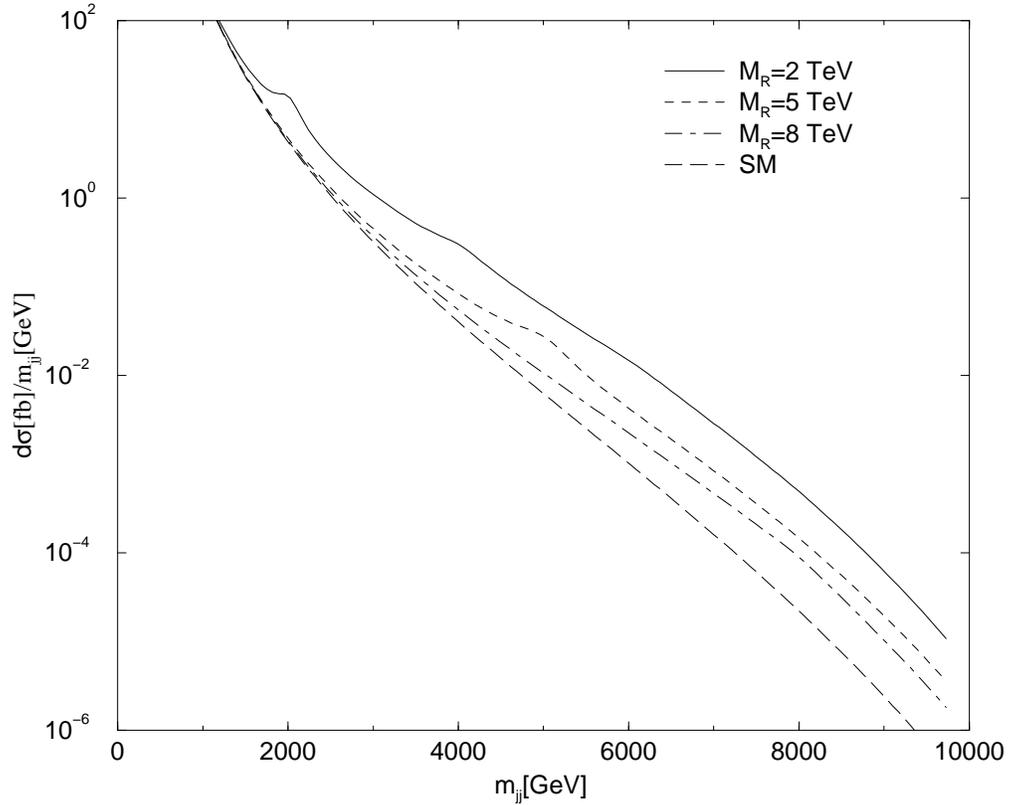}
\caption{The differential cross section $d\sigma/dm_{jj}$  
($j\neq t$) for dijet 
production in the process $pp\rightarrow jj+X$ including Kaluza-Klein  
gluon exchange
as a function of the dijet invariant mass $m_{jj}$ for the cases 
when the mass scale  $M_R$ of the compactified dimension is 2 TeV, 
5 TeV, 8 TeV, and for the SM. The cross section is evaluated at the leading
order. A rapidity cut $\eta<0.5$ is imposed for both jets.}
\label{fig4}
\end{center}
\end{figure}

\begin{figure}
\begin{center}
\includegraphics[angle=270,width=6.0in]{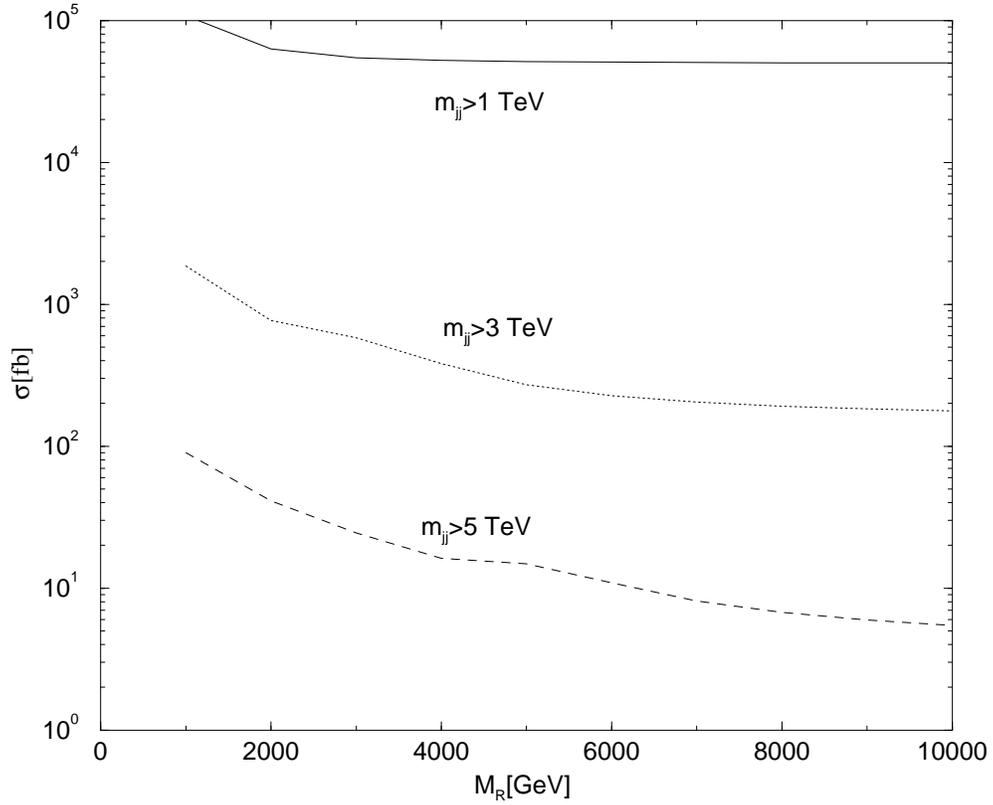}
\caption{Plot of the production cross section $pp\rightarrow jj+X$ ($j \neq t$)
as a function of the compactified dimension with jj invariant mass
cut of $m_{jj}>1 $ TeV,  $m_{jj}>3 $ TeV, and $m_{jj}>5 $ TeV.
A rapidity cut $\eta<0.5$ is imposed for both jets. }
\label{fig5}
\end{center}
\end{figure}

\begin{figure}
\begin{center}
\includegraphics[angle=270,width=6.0in]{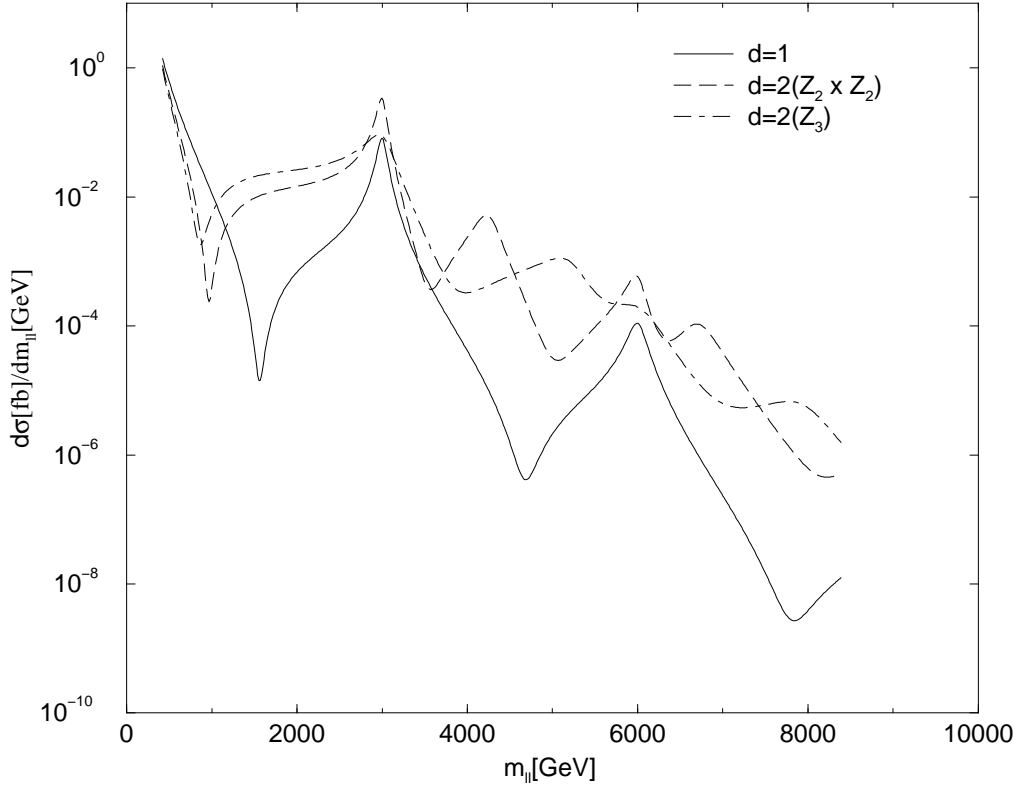}
\caption{Plot of the cross section for dilepton production as a function
of the dilepton invariant mass $m_{ll}$ for the case d=1 (solid),
d=2 with $Z_2\times Z_2$ orbifolding (dashed) and d=2 with $Z_3$ orbifolding
(dot-dashed) when the mass of the first Kaluza-Klein  excitation is taken to
be 3 TeV. }
\label{fig6}
\end{center}
\end{figure}

\end{document}